\documentclass[sigconf]{acmart}

\usepackage{amsfonts}
\usepackage{amsmath}
\usepackage{booktabs}
\usepackage{float}
\usepackage{graphicx}
\usepackage{multirow}
\usepackage{url}
\usepackage{xcolor}
\usepackage{xspace}
\usepackage{caption}
\usepackage{paralist}

\newcommand{\vc}[1]{\mathbf{#1}}
\newcommand{\mat}[1]{\mathbf{#1}}
\newcommand{\querymatrix}{\mat{X}}
\newcommand{\labelmatrix}{\mat{Y}}

\newcommand{\weightmatrixat}[1]{\mat{W}^{(#1)}}
\newcommand{\maskmatrix}{\mathcal{M}}
\newcommand{\maskmatrixat}[1]{\maskmatrix^{(#1)}}

\newcommand{\featuredim}{d}
\newcommand{\treedepth}{D}
\newcommand{\querycount}{n}
\newcommand{\labelcount}{L}
\newcommand{\clustercount}{K}
\newcommand{\query}{\vc{x}}
\newcommand{\product}{\vc{y}}
\newcommand{\prodemb}{\vc{z}}
\newcommand{\weight}{\vc{w}}
\newcommand{\weightfull}[2]{\vc{w}^{(#1)}_{#2}}
\newcommand{\Rn}[1]{\mathbb{R}^{#1}}

\newcommand{\labelspace}{\mathcal{Y}}
\newcommand{\bigO}{\mathcal{O}}
\newcommand{\clusterinst}{\tilde{\mathcal{Y}}}
\newcommand{\cluster}[2]{\mathcal{Y}^{(#1)}_{#2}}

\newcommand{\layer}{t}

\newcommand{\labelinst}{\ell}

\newcommand{\clusterind}{\mat{C}}
\newcommand{\clusterindat}[1]{\clusterind^{(#1)}}

\newcommand{\best}[1]{\bf #1}
\newcommand{\pecos}{PECOS\xspace}
\newcommand{\liblinear}{LIBLINEAR\xspace}
\newcommand{\xmc}{XMC\xspace}
\newcommand{\xmclong}{eXtreme Multi-label Classification \xspace}
\newcommand{\xlinear}{{X-Linear}\xspace}
\newcommand{\xtransformer}{{X-Transformer}\xspace}
\newcommand{\xrlinear}{{XR-Linear}\xspace}

\newcommand{\pifa}{{\sf PIFA}\xspace}
\newcommand{\tfidf}{TF-IDF\xspace}

\newcommand{\attentionxml}{AttentionXML\xspace}
\newcommand{\annexml}{AnnexML\xspace}
\newcommand{\bonsai}{Bonsai\xspace}
\newcommand{\discmec}{DiSMEC\xspace}

\newcommand{\parabel}{Parabel\xspace}

\newcommand{\ppdsparse}{PPD-Sparse\xspace}
\newcommand{\proxml}{ProXML\xspace}

\newcommand{\slice}{SLICE\xspace}

\newcommand{\xtext}{eXtremeText\xspace}
\newcommand{\xreg}{XReg\xspace}
\newcommand{\napkinxc}{NAPKINXC\xspace}
\newcommand{\lightxml}{LightXML\xspace}
\newcommand{\hnsw}{{HNSW}\xspace}
\newcommand{\scann}{{ScaNN}\xspace}
\newcommand{\knn}{{kNN}\xspace}
\newcommand{\nmslib}{{NMSLIB}\xspace}
\newcommand{\dssm}{{DSSM}\xspace}
\newcommand{\bmtf}{{Okapi-BM25}\xspace}
\AtBeginDocument{%
  \providecommand\BibTeX{{%
    \normalfont B\kern-0.5em{\scshape i\kern-0.25em b}\kern-0.8em\TeX}}}

\copyrightyear{2021}
\acmYear{2021}
\setcopyright{acmcopyright}
\acmConference[KDD '21]{Proceedings of the 27th ACM SIGKDD Conference on Knowledge Discovery and Data Mining}{August 14--18, 2021}{Virtual Event, Singapore}
\acmBooktitle{Proceedings of the 27th ACM SIGKDD Conference on Knowledge Discovery and Data Mining (KDD '21), August 14--18, 2021, Virtual Event, Singapore}
\acmPrice{15.00}
\acmDOI{10.1145/3447548.3467092}
\acmISBN{978-1-4503-8332-5/21/08}
\begin{document}
\fancyhead{}

\title{Extreme Multi-label Learning for \\ Semantic Matching in Product Search}


\author{Wei-Cheng Chang}
\affiliation{%
  	\institution{Amazon}
  	%\city{Palo Alto}
  	\country{USA}
}

\author{Daniel Jiang}
\affiliation{%
  	\institution{Amazon}
  	%\city{Palo Alto}
  	\country{USA}
}

\author{Hsiang-Fu Yu}
\affiliation{%
  	\institution{Amazon}
  	%\city{Palo Alto}
  	\country{USA}
}

\author{Choon-Hui Teo}
\affiliation{%
  	\institution{Amazon}
  	%\city{Palo Alto}
  	\country{USA}
}

\author{Jiong Zhang}
\affiliation{%
  	\institution{Amazon}
  	%\city{Palo Alto}
  	\country{USA}
}

\author{Kai Zhong}
\affiliation{%
  	\institution{Amazon}
  	%\city{Palo Alto}
  	\country{USA}
}

\author{Kedarnath Kolluri}
\affiliation{%
  	\institution{Amazon}
  	%\city{Palo Alto}
  	\country{USA}
}

\author{Qie Hu}
\affiliation{%
  	\institution{Amazon}
  	%\city{Palo Alto}
  	\country{USA}
}

\author{Nikhil Shandilya}
\affiliation{%
  	\institution{Amazon}
  	%\city{Palo Alto}
  	\country{USA}
}

\author{Vyacheslav Ievgrafov}
\affiliation{%
  	\institution{Amazon}
  	%\city{Palo Alto}
  	\country{USA}
}

\author{Japinder Singh}
\affiliation{%
  	\institution{Amazon}
  	%\city{Palo Alto}
  	\country{USA}
}

\author{Inderjit S. Dhillon}
\affiliation{%
  	\institution{Amazon}
  	%\city{Palo Alto}
  	\country{USA}
}

\renewcommand{\shortauthors}{Trovato and Tobin, et al.}

\begin{abstract}
	We consider the problem of semantic matching in product search: given a customer query,
	retrieve all semantically related products from a huge catalog of size $100$ million, or more.
	Because of large catalog spaces and real-time latency constraints,
	semantic matching algorithms not only desire high recall but also need to have low latency.
	Conventional lexical matching approaches (e.g., \bmtf)
	exploit inverted indices to achieve fast inference time,
	but fail to capture behavioral signals between queries and products.
	In contrast, embedding-based models learn semantic representations from customer behavior data,
	but the performance is often limited by shallow neural encoders due to latency constraints.
	Semantic product search can be viewed as an \xmclong (\xmc) problem,
	where customer queries are input instances and products are output labels.
	In this paper, we aim to improve semantic product search by using tree-based \xmc models
	where inference time complexity is logarithmic in the number of products.
	We consider hierarchical linear models with n-gram features for fast real-time inference.
	Quantitatively, our method maintains a low latency of $1.25$ milliseconds per query
	and achieves a $65\%$ improvement of Recall@100 ($60.9\%$ v.s. $36.8\%$)
	over a competing embedding-based \dssm model.
	Our model is robust to weight pruning with varying thresholds,
	which can flexibly meet different system requirements for online deployments.
	Qualitatively, our method can retrieve products that are
	complementary to existing product search system and add diversity to the match set.
\end{abstract}




\maketitle

\section{Introduction}

\begin{figure}[!ht]
	\centering
	\includegraphics[width=\linewidth]{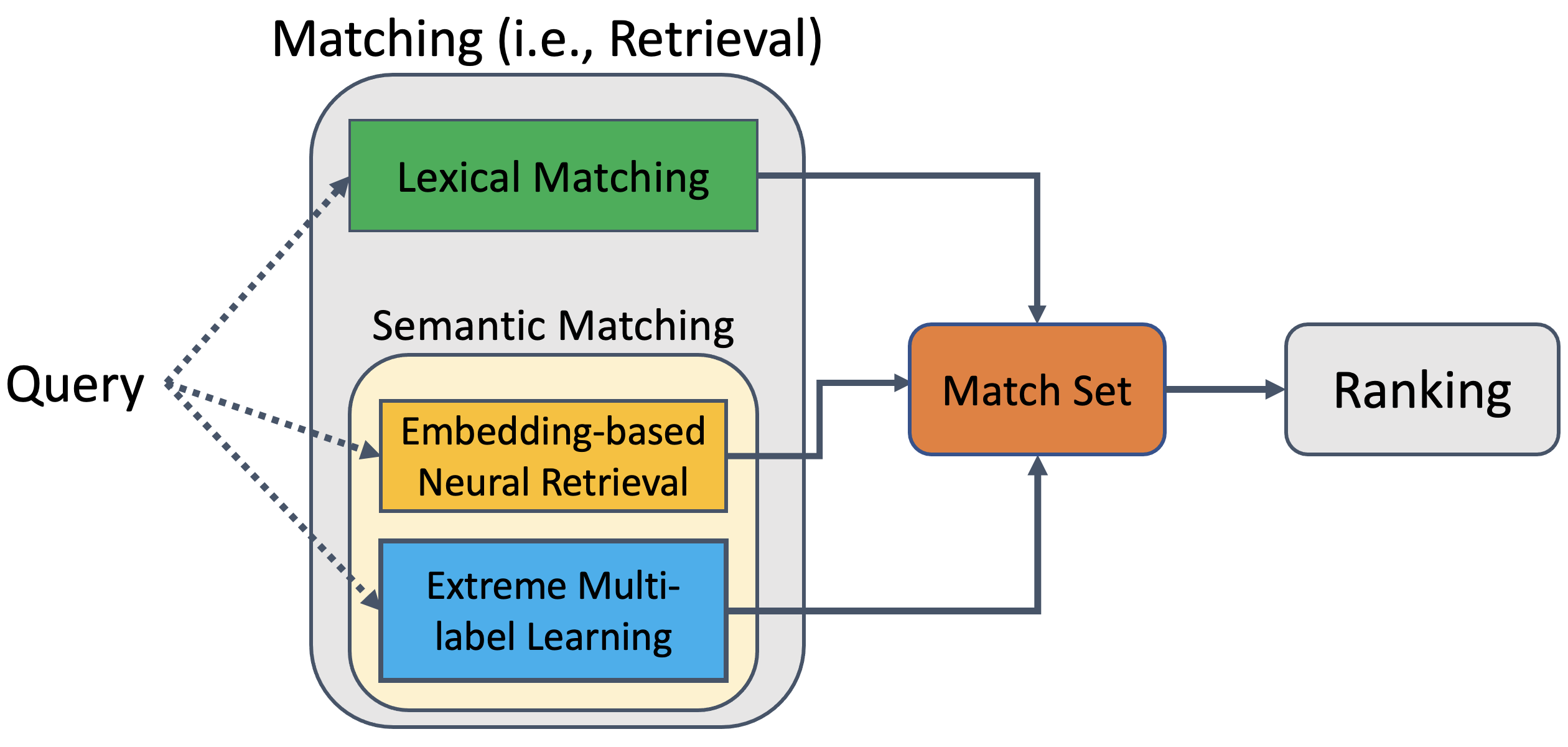}
	\caption{
		System architecture for augmenting match set with
		extreme multi-label learning models.
	}
	\vspace{-.5em}
	\label{fig:product-search}
	\vspace{-.5em}
\end{figure}

In general, product search consists of two building blocks:
the \textit{matching} stage, followed by the \textit{ranking} stage.
When a customer issues a query, the query is passed to several matching algorithms
to retrieve relevant products, resulting in a \textit{match set}.
The match set passes through stages of ranking, where top results from the previous
stage are re-ranked before the most relevant items are displayed to customers.
Figure~\ref{fig:product-search} outlines a product search system.

The matching (i.e., retrieval) phase is critical. Ideally,
the match set should have a high recall~\cite{manning08introduction},
containing as many relevant
and diverse products that match the customer intent as possible; otherwise,
many relevant products won't be considered in the final ranking phase.
On the other hand, the matching algorithm should be highly efficient
to meet real-time inference constraints:
retrieving a small subset of relevant products in time sublinear to
enormous catalog spaces, whose size is as large as 100 million or more.

% lexical matching
Matching algorithms can be put into two categories.
The first type is lexical matching approaches that score a query-product pair
by a weighted sum of overlapping keywords among the pair.
One representative example is \bmtf~\cite{robertson1994some,robertson2009probabilistic},
which remains state-of-the-art for decades and is still widely used in
many retrieval tasks such as open-domain question answering~\cite{lee2019latent,chang2020pretraining}
and passage/document retrieval~\cite{gao2020complementing,boytsov2020flexible}.
While the inference of \bmtf can be done efficiently using an inverted index~\cite{zobel2006inverted},
these index-based lexical matching approaches cannot capture semantic
and customer behavior signals (e.g., clicks, impressions, or purchases)
tailored from the product search system
and are fragile to morphological variants or spelling errors.

% embedding-based neural retrieval
The second option is embedding-based neural models
that learn semantic representations of queries and products based
on the customer behavior signals.
The similarity is measured by inner products or cosine distances
between the query and product embeddings.
To infer in real-time the match set of a novel query,
embedding-based neural models need to first
\textit{vectorize} the query tokens into the query embedding,
and then find its nearest neighbor products in the embedding space.
Finding nearest neighbors can be done efficiently via
approximate nearest neighbor search (ANN) methods
(e.g., \hnsw~\cite{malkov2020hnsw} or \scann~\cite{guo2020accelerating}).
Vectorizing query tokens into an embedding,
nevertheless, is often the inference bottleneck,
that depends on the complexity of neural network architectures.

% Embedding-based Inference example: {MLP vs BERT-based} + HNSW
Using BERT-based encoders~\cite{devlin2018bert} for query embeddings
is less practical because of the large latency in vectorizing queries.
Specifically, consider retrieving the match set of a query from
the output space of $100$ millions of products on a CPU machine.
In Table~\ref{tb:inference-toy},
we compare the latency of a BERT-based encoder ($12$ layers deep Transformer)
with a shallow \dssm~\cite{nigam2019semantic} encoder ($2$ layers MLP).
The inference latency of a BERT-based encoder is $50$ milliseconds per query
(ms$/$q) in total, where vectorization and ANN search needs $88\%$ and
$12\%$ of the time, respectively.
On the other hand, the latency of \dssm encoder is $3.13$ ms$/$q in total,
where vectorization only takes $33\%$ of the time.
\begin{table}[!ht]
	\centering
	\resizebox{\columnwidth}{!}{%
	\begin{tabular}{cr|cr}
		\toprule
		Vectorizer 							& latency (ms$/$q) 	& ANN & latency (ms$/$q) \\
		\midrule
		\dssm~\cite{nigam2019semantic}		&  1.00 	& \hnsw~\cite{malkov2020hnsw} 	& 2.13 \\
		BERT-based~\cite{devlin2018bert}	& 44.00 	& \hnsw~\cite{malkov2020hnsw} 	& $\approx 6.39$ \\
		\bottomrule
	\end{tabular}%
	}
	\caption{
		Comparing inference latency of BERT-based encoder with \dssm encoder on a CPU machine,
		with single-batch, single-thread settings.
		The latency of BERT-based encoder is taken from HuggingFace Transformers~\cite{wolf2020transformers,wolf2021benchmark}.
		BERT's ANN time is estimated by linear extrapolation on the dimensionality $d$,
		where BERT has $d=768$ and \dssm has $d=256$.
	}
	\vspace{-2em}
	\label{tb:inference-toy}
\end{table}
Because of real-time inference constraints, industrial product search engines
do not have the luxury to leverage state-of-the-art deep Transformer encoders
in embedding-based neural methods,
whereas shallow MLP encoders result in limited performance.

% Motivation of XMC
In this paper, we take another route to tackle semantic matching
by formulating it as an \xmclong (\xmc) problem,
where the goal is tagging an input instance (i.e., query) with
most relevant output labels (i.e., products).
\xmc approaches have received great attention recently
in both the academic and industrial community.
For instance, the deep learning \xmc model \xtransformer~\cite{chang2020xmctransformer,yu2020pecos}
achieves state-of-the-art performance on public academic benchmarks~\cite{bhatia2016xmc}.
Partition-based methods such as \parabel~\cite{prabhu2018parabel} and \xreg~\cite{prabhu2020extreme},
as another example, finds successful applications to dynamic search advertising in Bing.
In particular, tree-based partitioning \xmc models are a staple
of modern search engines and recommender systems due to their inference time
being sub-linear (i.e., logarithmic) to the enormous output space (e.g., $100$ million or more).

In this paper, we aim to answer the following research question:
\textit{Can we develop an effective tree-based \xmc approach for semantic matching in product search?}
Our goal is not to replace lexical-based matching methods (e.g., \bmtf).
Instead, we aim to complement/augment the match set with more diverse candidates
from the proposed tree-based \xmc approaches;
hence the final stage ranker can produce a good and diverse set of products
in response to search queries.

% Our contribution
We make the following contributions in this paper.
%\begin{itemize}[noitemsep,topsep=0pt,parsep=0pt,partopsep=0pt,leftmargin=*]
\begin{itemize}
	\item We apply the leading tree-based \xrlinear (\pecos) model~\cite{yu2020pecos} to semantic matching.
		To our knowledge, we are the first to successfully apply
		tree-based \xmc methods to an industrial scale product search system.
	\item We explore various $n$-gram \tfidf features for \xrlinear as
		vectorizing queries into sparse TF-IDF features is highly efficient for real-time inference.
	\item We study weight pruning of the \xrlinear model and demonstrate its robustness to different disk-space requirements.
	\item We present the trade-off between recall rates and real-time latency.
		Specifically, for beam size $b=15$, \xrlinear achieves Recall$@100$ of $60.9\%$ with a latency
		of $1.25$ ms$/$q; for beam size $b=50$, \xrlinear achieves Recall$@100$ of $65.7\%$
		with a latency of $3.48$ ms$/$q.
		In contrast, \dssm only has Recall$@100$ of $36.2\%$ with a latency of $3.13$ ms$/$q.
\end{itemize}
The implementation of \xrlinear (\pecos) is publicly available at
\url{https://github.com/amzn/pecos}.
%when conducting real-time inference on an industrial-scale product search problem
%whose product space is at the scale of $100$ millions or more.

\section{Related Work}

\subsection{Semantic Product Search Systems}
Two-tower models (a.k.a., dual encoders, Siamese networks) are arguably
one of the most popular embedding-based neural models
used in passage or document retrieval~\cite{nguyen2016ms,xiong2020approximate},
dialogue systems~\cite{mazare2018training,henderson2019training},
open-domain question answering~\cite{lee2019latent,guu2020realm,karpukhin2020dense},
and recommender systems~\cite{covington2016deep,yi2019sampling,yang2020mixed}.
It is not surprising that two-tower models with ResNet~\cite{he2016deep} encoder
and deep Transformer encoders~\cite{vaswani2017attention}
achieve state-of-the-art in most computer vision~\cite{kuznetsova2020open}
and textual-domain retrieval benchmarks~\cite{craswell2020overview}, respectively.
The more complex the encoder architecture is, nevertheless,
the less applicable the deep two-tower models are for industrial product search systems:
very few of them can meet the low real-time latency constraint (e.g., 5 ms/q)
due to the vectorization bottleneck of query tokens.

One of the exceptions is Deep Semantic Search Model (\dssm)~\cite{nigam2019semantic},
a variant of two-tower retrieval models with shallow multi-layer perceptron (MLP) layers.
\dssm is tailored for industrial-scale semantic product search,
and was A/B tested online on Amazon Search Engine~\cite{nigam2019semantic}.
Another example is two-tower recommender models for Youtube~\cite{covington2016deep}
and Google Play~\cite{yang2020mixed}, where they also embrace
shallow MLP encoders for fast vectorization of novel queries
to meet online latency constraints.
Finally, \cite{song2020large} leverages distributed GPU training and KNN softmax,
scaling two-tower models for the Alibaba Retail Product Dataset with $100$ million products.

\subsection{\xmclong (\xmc)}
To overcome computational issues,
most existing \xmc algorithms with textual inputs use sparse \tfidf features
and leverage different partitioning techniques on the label space to reduce complexity.

\paragraph{\bf Sparse linear models}
Linear one-versus-rest~(OVR) methods such as
\discmec~\citep{babbar2017dismec}, \proxml~\citep{babbar2019data},
\ppdsparse~\citep{yen2016pd,yen2017ppdsparse} explore
parallelism to speed up the algorithm and reduce the model size by truncating
model weights to encourage sparsity.
Linear OVR classifiers are also building blocks for many other \xmc approaches,
such as
\parabel~\citep{prabhu2018parabel},
\slice~\citep{jain2019slice},
\xrlinear(\pecos)~\citep{yu2020pecos},
to name just a few.
However, naive OVR methods are not applicable to semantic product search
because their inference time complexity is still linear in the output space.

\paragraph{\bf Partition-based methods}
The efficiency and scalability of sparse linear models can be further improved
by incorporating different partitioning techniques on the label spaces.
For instance, \parabel~\citep{prabhu2018parabel} partitions the labels through a
balanced 2-means label tree using label features constructed from the instances.
Other approaches attempt to improve on \parabel, for instance,
\xtext~\citep{wydmuch2018no}, \bonsai~\citep{khandagale2019bonsai},
\napkinxc~\citep{jasinska2020probabilistic},
\xreg~\citep{prabhu2020extreme}, and
~\pecos~\citep{yu2020pecos}.
In particular, the \pecos framework~\cite{yu2020pecos}
allows different indexing and matching methods for \xmc problems,
resulting in two realizations, \xrlinear and \xtransformer.
Also note that tree-based methods with neural encoders such as
\attentionxml~\citep{you2019attentionxml},
\xtransformer(\pecos)~\citep{chang2020xmctransformer,yu2020pecos},
and \lightxml~\citep{jiang2021lightxml},
are mostly the state-of-the-art results on \xmc benchmarks,
at the cost of longer training time and expensive inference.
To sum up, \textit{tree-based sparse linear methods} are more suitable
for the industrial-scale semantic matching problems due to
their sub-linear inference time and fast vectorization of the query tokens.

\paragraph{\bf Graph-based methods}
\slice~\citep{jain2019slice} and \annexml~\citep{tagami2017annexml}
building an approximate nearest neighbor (ANN) graph to index the large output space.
For a given instance, the relevant labels can be found quickly via ANN search.
\slice then trains linear OVR classifiers with negative samples induced from ANN.
While the inference time complexity of advanced ANN algorithms
(e.g., \hnsw~\cite{malkov2020hnsw} or \scann~\cite{guo2020accelerating})
is sub-linear, ANN methods typically work better on low-dimensional dense embeddings.
Therefore, the inference latency of \slice and \annexml still hinges
on the vectorization time of pre-trained embeddings.

\section{\xrlinear(\pecos): a tree-based \xmc method for semantic matching}
%In this work, we apply the leading tree-based \xmc method \xrlinear~\cite{yu2020pecos}
%to semantic matching for product search systems.
%We first discuss learning/estimating model of \xrlinear,
%followed by pruning/inference of \xrlinear for semantic product search.

\paragraph{\bf Overview}
Tree-based \xmc models for semantic matching can be characterized as follows:
given a vectorized query $\query \in \Rn{d}$ and a set of
labels (i.e., products) $\labelspace=\{1, \ldots, \labelinst, \ldots, \labelcount\}$,
produce a tree-based model that retrieves top $k$ most relevant products in $\labelspace$.
The model parameters are estimated from the training dataset
$\{(\query_i, \product_i): i=1,\ldots,n\}$ where $\product_i \in \{0,1\}^{\labelcount}$
denotes the relevant labels for this query from the output label space $\labelspace$.

\begin{figure}[!ht]
	\centering
	\includegraphics[width=0.95\linewidth]{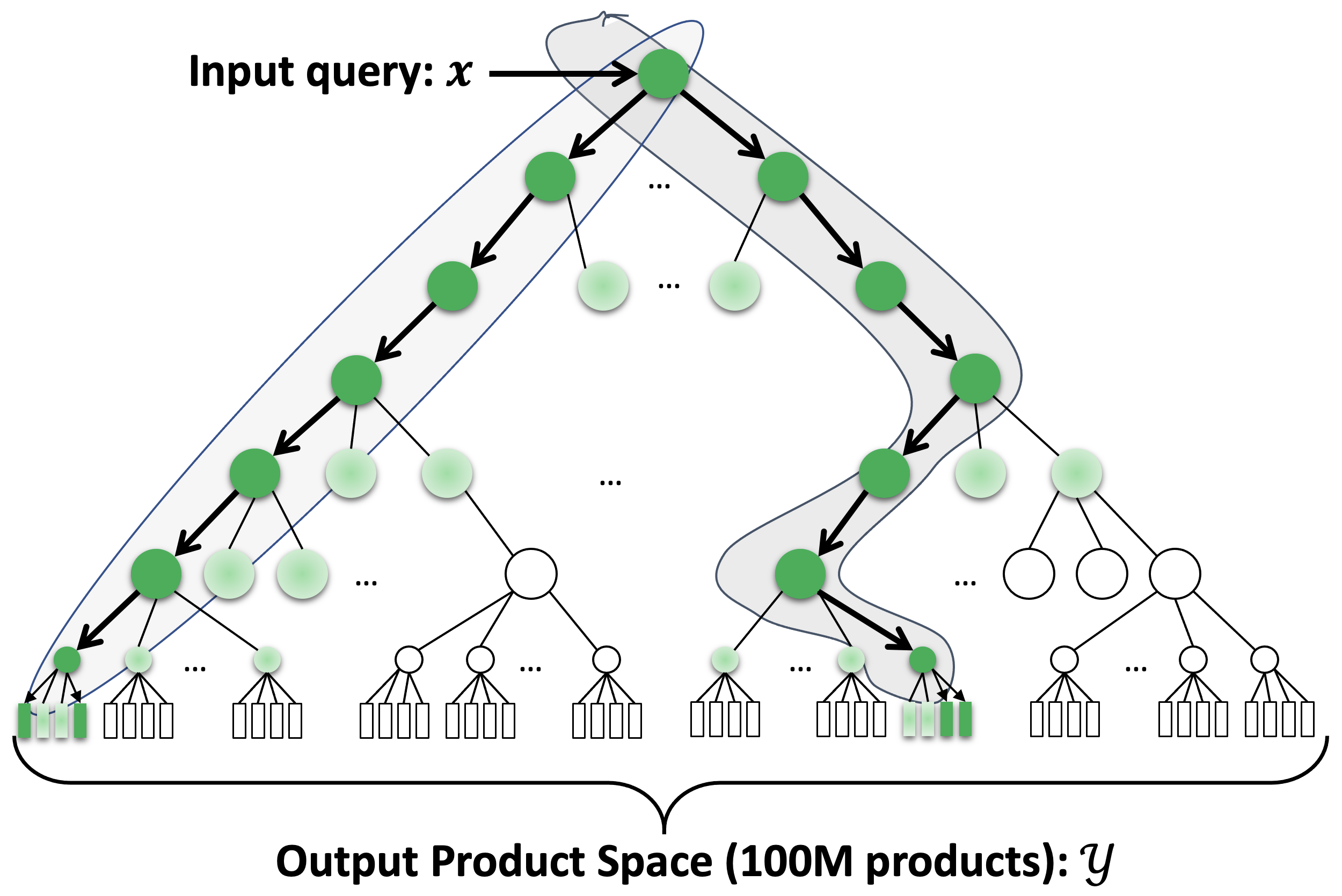}
	\caption{
		Illustration of inference of \xrlinear(\pecos)~\cite{yu2020pecos} using beam search with beam width $b=2$
		to retrieve $4$ relevant products for the given input query $\query$.
	}
	\label{fig:inference}
\end{figure}

% clustering
There are many tree-based linear methods
~\cite{prabhu2018parabel,wydmuch2018no,khandagale2019bonsai,jasinska2020probabilistic,prabhu2020extreme,yu2020pecos},
and we consider the \xrlinear model within \pecos~\cite{yu2020pecos} in this paper,
due to its flexibility and scalability to large output spaces.
The \xrlinear model partitions the enormous label space $\labelspace$
with hierarchical clustering to form a hierarchical label tree.
The $j$th label cluster at depth $\layer$ is denoted by $\cluster{\layer}{j}$.
The leaves of the tree are the individual labels (i.e., products) of $\labelspace$.
See Figure~\ref{fig:inference} for an illustration
of the tree structure and inference with beam search.

% OVR classifiers and predictions
Each layer of the \xrlinear model has a \textit{linear} OVR classifier that scores the
relevance of a cluster $\cluster{\layer}{j}$ to a query $\query$.
Specifically, the unconditional relevance score of a cluster $\cluster{\layer}{j}$
is
\begin{equation}
	f(\query, \cluster{\layer}{j})
	= \sigma ( \query^{\top} \weightfull{\layer}{j} ),
\end{equation}
where $\sigma$ is an activation function and
$\weightfull{\layer}{j} \in \Rn{d}$ are model parameters of the $j$-th node at $\layer$-th layer.

Practically, the weight vectors $\weightfull{\layer}{j}$ for each layer $\layer$
are stored in a $d \times \clustercount_\layer$ weight matrix
\begin{equation}
	\weightmatrixat{\layer} =
	\left[\begin{array}{cccc}
		\weightfull{\layer}{1} & \weightfull{\layer}{2} & ... & \weightfull{\layer}{\clustercount_\layer}
	\end{array}\right] \,,
\end{equation}
where $\clustercount_\layer$ denotes the number of clusters at layer $\layer$,
such that $\clustercount_\treedepth = \labelcount$ and $\clustercount_0 = 1$.
In addition, the tree topology at layer $\layer$ is represented by a cluster indicator matrix
$\clusterindat{\layer} \in \{0, 1\}^{\clustercount_{\layer + 1} \times \clustercount_\layer}$.
Next, we discuss how to construct the cluster indicator matrix $\clusterindat{\layer}$
and learn the model weight matrix $\weightmatrixat{\layer}$.

\subsection{Hierarchical Label Tree}
\label{sec:label-tree}

\begin{figure}[!ht]
	\centering
	\includegraphics[width=0.95\linewidth]{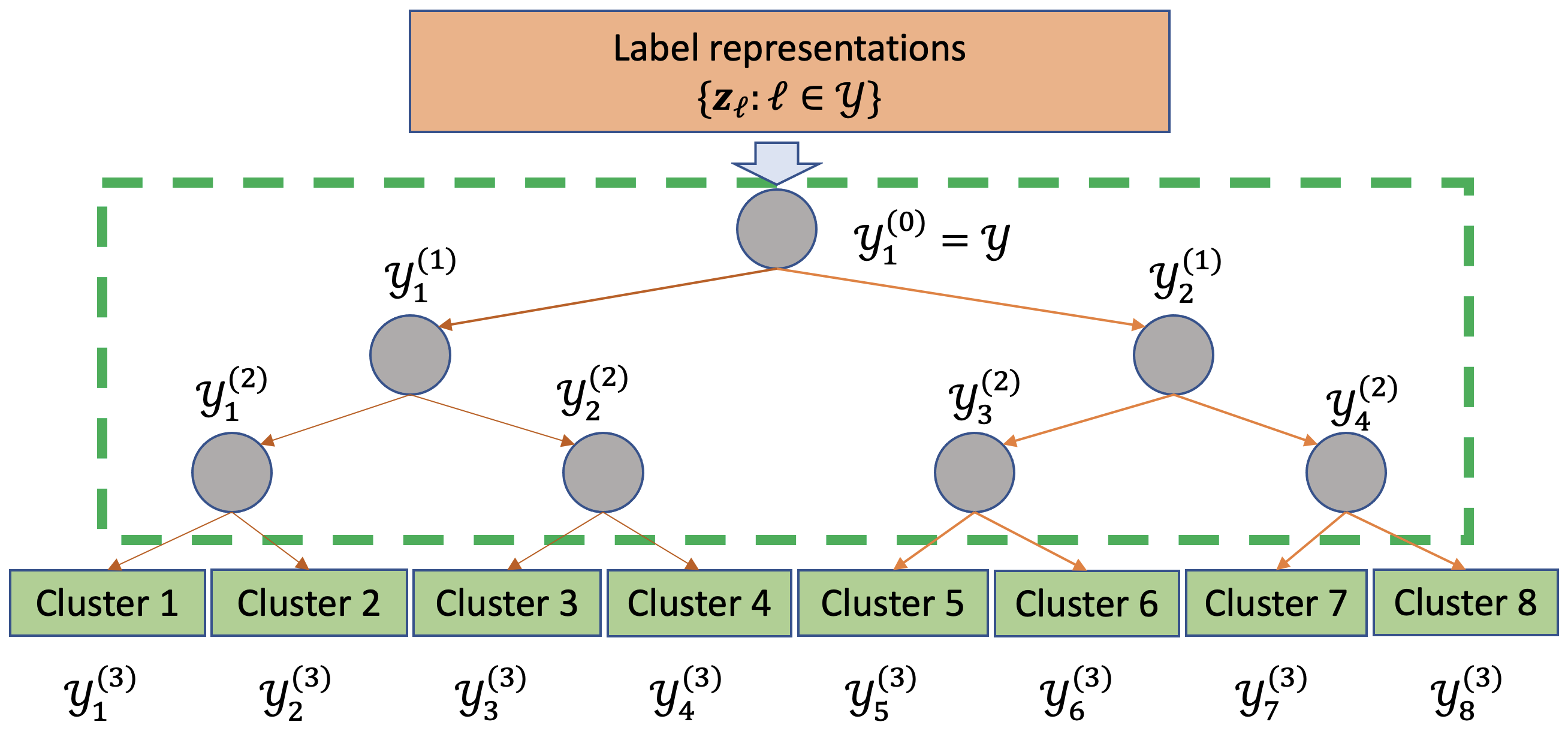}
	\vspace{-.5em}
	\caption{
		An illustration of hierarchical label clustering~\cite{yu2020pecos}
		with recursive $B$-ary partitions with $B=2$.
	}
	\vspace{-.5em}
	\label{fig:clustering}
\end{figure}

\paragraph{\bf Tree Construction}
Given semantic label representations $\{\prodemb_\labelinst: \labelspace\}$,
the \xrlinear model constructs a hierarchical label tree via $B$ array partitioning
(i.e., clustering) in a top-down fashion, where the number of clusters at layer
$\layer=1, \ldots, \treedepth$ are
\begin{equation}
	\clustercount_\treedepth
	= \labelcount, \clustercount_{\treedepth-1} = B^{\treedepth-1}, \ldots, \clustercount_1 = B^1.
\end{equation}
Starting from the node $k$ of layer $\layer-1$ is a cluster $\cluster{\layer}{k}$ containing
labels assigned to this cluster. For example, $\cluster{0}{1}=\labelspace$ is the root node
clustering including all the labels $\{\labelinst: 1,\ldots,\labelcount\}$.
To proceed from layer $\layer - 1$ to layer $\layer$,
we perform Spherical $B$-Means~\cite{dhillon2001concept} clustering
to partition $\cluster{\layer-1}{k}$ into $B$ clusters to form
its $B$ child nodes $\{ \cluster{\layer}{B(k-1)+j} : j = 1, \ldots, B \}$.
Applying this for all parent nodes $k=1,\ldots, B^{\layer-1}$ at layer $\layer-1$,
the cluster indicator matrix $\clusterindat{\layer} \in \{0,1\}^{\clustercount_\layer \times \clustercount_{\layer-1}}$
at layer $\layer$ is represented by
\begin{equation}
	\Big( \clusterindat{\layer} \Big)_{j,k} =
	\begin{cases}
		1, &\text{ if } \cluster{\layer}{j} \text{ is a child of } \cluster{\layer-1}{k}, \\
		0, &\text{ otherwise}.
	\end{cases}
\end{equation}
An illustration with $B=2$ is given in Figure~\ref{fig:clustering}.

\paragraph{\bf Label Representations}
For semantic matching problem, we present two ways to construct label embeddings,
depending on the quality of product information (i.e., titles and descriptions).
If product information such as titles and descriptions are noisy or missing,
each label is represented by aggregating query feature vectors from positive instances (namely, \pifa):
\begin{equation}
	\prodemb_{\labelinst}^{\pifa}
	= \frac{\vc{v}_\labelinst}{\|\vc{v}_\labelinst\|}, \text{ where } \vc{v}_\labelinst
	= \sum_{i=1}^n \labelmatrix_{i \labelinst} \query_i,
	\quad \labelinst \in \labelspace.
\end{equation}
Otherwise, labels with rich textual information can be represented by
featurization (e.g., $n$-gram \tfidf or some pre-trained neural embeddings)
of its titles and descriptions.

\subsection{Sparse Linear Matcher}
\label{sec:model-learning}

Given the query feature matrix $\querymatrix \in \Rn{n \times d}$,
the label matrix $\labelmatrix \in \{0,1\}^{n \times \labelcount}$,
and cluster indicator matrices
$\{ \clusterindat{\layer} \in \{0,1\}^{\clustercount_t \times \clustercount_{t-1}} : \clustercount_0=1, \clustercount_\treedepth=\labelcount, t=1, \ldots, \treedepth\}$,
we are ready to learn the model parameters $\weightmatrixat{\layer}$
for layer $\layer=1,\ldots,\treedepth$ in a top-down fashion.
For example, at the bottom layer $t=\treedepth$, it corresponds
to the original \xmc problem on the given training data
$\{\querymatrix, \labelmatrix^\treedepth = \labelmatrix \}$.
For layer $t=\treedepth-1$, we construct
the \xmc problem with the induced dataset
$\{\querymatrix, \labelmatrix^{\treedepth - 1} = \text{binarize}(\labelmatrix^\treedepth \clusterindat{\treedepth}) \}$.
Similar recursion is applied to all other intermediate layers.

\paragraph{\bf Learning OVR Classifiers}
At layer $\layer$ with the corresponding \xmc dataset
$\{\querymatrix, \labelmatrix^\layer \}$,
the \xrlinear model considers point-wise loss to learn the
parameter matrix $\weightmatrixat{\layer} \in \Rn{\featuredim \times \clustercount_\layer}$
independently for each column $\labelinst$, $1\le \labelinst \le \clustercount_\layer$.
Column $\labelinst$ of $\weightmatrixat{\layer}$ is found by solving a binary classification sub-problem:
\begin{equation}
	\weightfull{\layer}{\labelinst}
	= \arg\min_{\weight} \sum_{i: \maskmatrixat{\layer}_{i, \vc{c}_{\labelinst}} \neq 0}
		\mathcal{L}(\labelmatrix_{i\labelinst}^{\layer}, \weight^{\top} \query_i)
		+ \frac{\lambda}{2} \| \weight \|_2^2,
		\quad \labelinst = 1, \ldots, \clustercount_\layer,
	\label{eq:binary-classifier}
\end{equation}
where $\lambda$ is the regularization coefficient,
$\mathcal{L}(\cdot,\cdot)$ is the point-wise loss function (e.g., squared hinge loss),
$\maskmatrixat{\layer} \in \{0,1\}^{\querycount \times \clustercount_{\layer-1}}$
is a \textit{negative sampling matrix} at layer $\layer$, and
$\vc{c}_{\labelinst} = \vc{c}_{\labelinst}^{\layer} \in \{ 1, \ldots, \clustercount_{\layer-1} \}$
is the cluster index of the $\labelinst$-th label (cluster) at the $\layer$-th layer.
Crucially, the negative sampling matrix $\maskmatrixat{\layer}$ not only finds
hard negative instances (queries) for stronger learning signals, but also significantly
reduces the size of the active instance set for faster convergence.

In practice, we solve the binary classification sub-problem ~\eqref{eq:binary-classifier}
via a state-of-the-art efficient solver \liblinear~\cite{fan2008liblinear}.
Furthermore, $\clustercount_\layer$ independent sub-problems at layer $\layer$
can be computed in an embarrassingly parallel manner to fully utilize the
multi-core CPU design in modern hardware.

\subsection{Inference}
\label{sec:model-inference}

\paragraph{\bf Weight Pruning}
The \xrlinear model is composed of $\treedepth$ OVR linear classifiers
$f^{(\layer)}(\query,\labelinst)$ parametrized by matrices
$\weightmatrixat{\layer} \in \Rn{\featuredim \times \clustercount_\layer}$,
$1\le \layer \le \treedepth$
Naively storing the dense parameter matrices is not feasible.
To overcome a prohibitive model size, we apply entry-wise weight pruning
to sparsify $\weightmatrixat{\layer}$~\cite{babbar2017dismec,prabhu2018parabel}.
Specifically, after solving $\weightfull{\layer}{\labelinst}$
for each binary classification sub-problem,
we perform a hard thresholding operation to truncate parameters with
magnitude smaller than a pre-specified value $\epsilon \geq 0$ to zero:
\begin{equation}
	\Big( \weightmatrixat{\layer} \Big)_{ij} =
	\begin{cases}
		\weightmatrixat{\layer}_{ij}, &\text{ if } |\weightmatrixat{\layer}_{ij}| > \epsilon, \\
		0, &\text{ otherwise}.
	\end{cases}
\end{equation}
We set the pruning threshold $\epsilon$ such that the parameter matrices can be
stored in the main memory for real-time inference.

\paragraph{\bf Beam Search}
The relevance score of a query-cluster pair $(\query,\clusterinst)$
is defined to be the aggregation of all ancestor nodes in the tree:
\begin{equation}
	f(\query, \clusterinst)
	= \prod_{\weight \in \mathcal{A}(\clusterinst)} \sigma \left( \weight \cdot \query \right) \,,
\end{equation}
where $\mathcal{A}(\clusterinst) = \{ \weightfull{\layer}{i} \mid \clusterinst \subset \cluster{\layer}{i}, \layer \neq 0 \}$
denotes the weight vectors of all ancestors of $\clusterinst$ in the tree, including $\clusterinst$ and disregarding the root.
Naturally, this definition extends all the way to the individual labels $\labelinst \in \labelspace$ at the bottom of the tree.

In practice, exact inference is typically intractable,
as it requires searching the entire label tree.
To circumvent this issue,
\xrlinear use a greedy \textit{beam search} of the tree as an approximation.
For a query $\query$, this approach discards any clusters at a given level
that do not fall into the top $b$ most relevant clusters, where $b$ is the
beam search width.
The inference time complexity of \xrlinear with beam search~\cite{yu2020pecos} is
\begin{equation}
	\sum_{\layer=1}^{\treedepth} \bigO
		\Big( b \times \frac{\clustercount_\layer}{\clustercount_{\layer-1}} \times T_f \big)
	= \bigO \Big( \treedepth \times b \times \max\big( B, \frac{\labelcount}{B^{\treedepth-1}} \big) \times T_f \Big),
\end{equation}
where $T_f$ is the time to compute relevance score $f^{(\layer)}(\query, \labelinst)$.
We see that if the tree depth $D=\bigO(\log_B \labelcount)$ and
max leaf size $\labelcount/B^{\treedepth-1}$ is a small constant such as $100$,
the overall inference time complexity for \xrlinear is
\begin{equation}
	\bigO ( b \times T_f \times \log \labelcount ),
\end{equation}
which is \textit{logarithmic} in the size of the original output space.

\subsection{Tokenization and Featurization}
To pre-process the input query or product title, we apply simple normalization
(e.g., lower case, remove basic punctuations) and use white space
to tokenize the query string into a sequence of smaller components
such as words, phrases, or characters.
We combine word unigrams, word bigrams, and character trigrams
into a bag of $n$-grams feature, and construct sparse high-dimensional
\tfidf features as the vectorization of the query.

\paragraph{\bf Word $n$-grams}
The basic form of tokenization is a word unigram. For example,
the word unigrams of "artistic iphone 6s case" are
["artistic", "iphone", "6s", "case"].
However, word unigrams lose word ordering information,
which leads us to use higher order $n$-grams, such as bigrams.
For example, the word bigram of the same query becomes
["artistic\#iphone", "iphone\#6s", "6s\#case"].
These $n$-grams capture phrase-level information which
helps the model to better infer the customer intent for search.

\paragraph{\bf Character Trigrams}
The string is broken into a list of all three-character sequences,
and we only consider the trigrams within \textit{word boundary}.
For example, the character trigrams of the query "artistic iphone 6s case"
are ["\#ar", "art", "rti", "tis", "ist", "sti", "tic", "ic\#", "\#ip", "iph", "phi", "hon", "one", "\#6s", "6s\#", "\#ca", "cas", "ase", "se\#"].
Character trigrams are robust to typos ("iphone" and "iphonr") and can handle
compound words ("amazontv" and "firetvstick") more naturally. Another advantage
for product search is the ability to capture similarity of model parts and sizes.

%%%%%%%%%%%%%%%%%%%%%%%%%%%%%%%%
% 100m-asin offline experiment
%%%%%%%%%%%%%%%%%%%%%%%%%%%%%%%%
\section{Experimental Results}

\subsection{Data Preprocessing}
We sample over $1$ billion \textit{positive} query-product pairs as the data set,
which covers over $240$ million queries and $100$ million products.
A \textit{positive} pair means aggregated counts of clicks or purchases are above a threshold.
While each query-product pair may have a real-valued weight representing the aggregated counts,
we do not currently leverage this information and simply treat the training signals as binary feedback.
We split the train/test set by time horizon where
we use $12$ months of search logs as the training set and the trailing $1$ month as the offline evaluation test set.
Approximately 12\% of products in the test set are unseen in the training set,
which are also known as cold-start products.

%Regarding the statistics of training set,
%the average number of products per query is $20.4$ and
%the average number of queries per product is $49.0$.
For the \xrlinear model and \bmtf,
we consider white space tokenization for both query text and product title
to vectorize the query feature and product features with $n$-gram \tfidf.
The training signals are extremely sparse, and the resulting
query feature matrix has an average of $27.3$ non zero elements per query
and the product feature matrix has an average of $142.1$ non zero element per product,
with the dimensionality $\featuredim=4.2M$.
For the \dssm model, both the query feature matrix and
product feature matrix are dense low-dimensional (i.e., 256) embeddings,
which are more suitable for \hnsw inference.
%The average number of tokens per query is $4.3$, and
%the average number of tokens per product title is $16.8$.
%Note that \xmc-based methods only leverage the textual information
%of query in the featurization step, while embedding-based retrieval
%approaches leverage the textual information of both query and product
%when learning the representations.

\paragraph{\bf Featurization}
For the proposed \xrlinear method, we consider $n$-gram sparse \tfidf features
for the input queries and conduct $\ell_2$ normalization of each query vector.
The vocabulary size is $4.2$ million,
which is determined by the most frequent tokens of word and character level $n$-grams.
In particular, there are $1$ million word unigrams,
$3$ million word bigrams, and $200$ thousand character trigrams.
Note that we consider characters within each word boundary when building
the character level $n$-grams vocabulary.
The motivation of using character trigrams is to better capture
specific model types, certain brand names, and typos.

For out-of-vocabulary (OOV) tokens,
we map those unseen tokens into a shared unique token id.
It is possible to construct larger bins with hashing tricks
~\cite{weinberger2009feature,joulin2017bag,nigam2019semantic}
that potentially handle the OOV tokens better.
We leave the hashing tricks as future work.

\subsection{Experimental Setup}

\paragraph{\bf Evaluation Protocol}
We sample $100,000$ queries as the test set, which comprises
$182,000$ query-product pairs with purchase counts being at least one.
Given a query in the retrieval stage, semantic matching algorithms
need to find the top $100$ relevant products from the catalog space
consisting of $100$ million candidate products.

We measure the performance with recall metrics, which are widely-used
in retrieval~\cite{chang2020pretraining,xiong2020approximate}
and \xmc tasks~\cite{prabhu2018parabel,reddi2018stochastic,jain2019slice}.
Specifically, for a predicted score vector $\hat{\vc{y}} \in \mathbb{R}^L$
and a ground truth label vector $\vc{y} \in \{0,1\}^L$, Recall$@k$ $(k=10,50,100)$
is defined as
\begin{equation*}
  \text{Recall}@k = \frac{\ell}{|\vc{y}|} \sum_{\ell \in \text{top}_k(\hat{\vc{y}})} \vc{y}_{\ell},
\end{equation*}
where $\text{top}_{k}(\hat{\vc{y}})$ denotes the labels with the largest $k$
predict values.

\paragraph{\bf Comparing Methods and Hyper-parameters}
We compare \xrlinear with two representative retrieval approaches
and describe the hyper-parameter settings.
\begin{itemize}
	\item \xrlinear:
		the proposed tree-based \xmc model, where
		the implementation is from \pecos~\cite{yu2020pecos}\footnote{\url{https://github.com/amzn/pecos}}.
		We use \pifa to construct semantic product embeddings
		and build the label tree via hierarchical K-means,
		where the branching factor is $B=32$, the tree depth is $D=5$, and
		the maximum number of labels in each cluster is $\labelcount/B^{\treedepth-1} = 100$.
		The negative sampling strategy is Teacher Forcing Negatives (TFN)~\cite{yu2020pecos},
		the transformation predictive function is $\mathcal{L}_3$ hinge,
		and default beam size is $b=10$. Unless otherwise specified,
		we follow the default hyper-parameter settings of \pecos.
	\item \dssm~\cite{nigam2019semantic}:
		the leading neural embedding-based retrieval model
		in semantic product search applications.
		We take a saved checkpoint \dssm model from ~\cite{nigam2019semantic},
		and generate query and product embeddings on our dataset for evaluation.
		For fast real-time inference, we use the state-of-the-art
		approximate nearest neighbor (ANN) search algorithm
		\hnsw~\cite{malkov2020hnsw} to retrieve relevant products efficiently.
		Specifically, we use the implementation of
		\nmslib~\cite{boytsov2013engineering}\footnote{\url{https://github.com/nmslib/nmslib}}.
		Regarding the hyper-parameter of \hnsw, we set $M_0=32$, $efC=300$,
		and vary the beam search width $efS=\{10, 50, 100, 250, 500\}$
		to see trade-off between inference latency and retrieval quality.
	\item \bmtf~\cite{robertson2009probabilistic}:
		the lexical matching baseline widely-used in retrieval tasks.
		We use the implementation of~\cite{trotman2014improvements}\footnote{\url{https://github.com/dorianbrown/rank_bm25}}
		where hyper-parameters are set to $k_1=0.5$ and $b=0.45$.
\end{itemize}

\paragraph{\bf Computing Environments}
We run \xrlinear and \bmtf on a
x1.32xlarge AWS instance ($128$ CPUs and $1952$ GB RAM).
For the embedding-based model \dssm,
the query and product embeddings are generated
from a pre-trained \dssm model~\cite{nigam2019semantic}
on a p3.16xlarge AWS instance (8 Nvidia V100 GPUs and $128$ GB RAM).
The inference of \dssm leverages the ANN algorithm \hnsw,
which is conducted on the same x1.32xlarge AWS instance
for fair comparison of inference latency.

\subsection{Main Results}

Comparison of \xrlinear(\pecos) with two representative retrieval approaches
(i.e., the neural embedding-based model \dssm and the lexical matching method \bmtf)
is presented in Table~\ref{tb:main-results}.
For \xrlinear, we study different $n$-gram TF-IDF feature realizations.
Specifically, we consider four configurations:
1) word-level unigrams with $3$ million features;
2) word-level unigrams and bigrams with $3$ million features;
3) word-level unigrams and bigrams with $5.6$ million features;
4) word-level unigrams and bigrams, plus character-level trigrams,
total of $4.2$ million features.
Configuration 3) has a larger vocabulary size compared to 2) and 4)
to examine the effect of character trigrams.
%\textcolor{red}{isd: maybe say why there are 5.6 million features in 3) even though it does not have trigrams?}

From $n$-gram TF-IDF configuration 1) to 3),
we observe that adding bigram features increases the Recall$@100$
from $54.86\%$ to $58.88\%$, a significant improvement over unigram features alone.
However, this also comes at the cost of a $1.57$x larger model size on disk.
We will discuss the trade-off between model size and inference latency in Section
~\ref{sec:exp-latency}.
Next, from configuration 3) to 4),
adding character trigrams further improves the Recall$@100$
from $58.88\%$ to $60.30\%$, with a slightly larger model size.
This empirical result suggests character trigrams can better handle
typos and compound words, which are common patterns in real-world product search system.
Similar observations are also found in ~\cite{huang2013learning,nigam2019semantic}.

All $n$-gram TF-IDF configurations of \xrlinear
achieve higher Recall$@k$ compared to the competitive
neural retrieval model \dssm with approximate nearest neighbor
search algorithm \hnsw.
For sanity check, we also conduct exact kNN inference
on the \dssm embeddings, and the conclusion remains the same.
Finally, the lexical matching method \bmtf performs poorly on this
semantic product search dataset, which may be due to the fact that
purchase signals are not well correlated with overlapping token statistics.
This indicates another unique difference between semantic product search
problem and the conventional web retrieval problem where \bmtf variants
are often a strong baseline~\cite{chen2017reading,lee2019latent,gao2020complementing}.

\paragraph{\bf Training Cost on AWS}
We compare the AWS instance cost of \xrlinear(\pecos) and \dssm, where the former
is trained on a x1.32xlarge CPU instance ($\$13.3$ per hour) and the latter
is trained on a p3.8xlarge GPU instance ($\$24.5$ per hour).
From Table~\ref{tb:main-results}, the training of \xrlinear takes $42.5$
hours (i.e., includes vectorization, clustering, matching),
which amounts to $\$567$ USD.
In contrast, the training of \dssm takes $260.0$ hours
(i.e., training \dssm with \textit{early stopping}, building \hnsw indexer),
which amounts to $\$6365$ USD.
In other words, the proposed \xrlinear enjoys a $10x$ smaller training cost
compared to \dssm, and thus offers a more frugal solution for
large-scale applications such as product search systems.

\begin{table*}[!ht]
	\centering
	\resizebox{0.975\textwidth}{!}{%
	\begin{tabular}{l|ccc|rr}
		\toprule
		Methods & Recall$@10$ & Recall$@50$ & Recall$@100$ & Training Time (hr) & Model Size (GB) \\
		\toprule
		\xrlinear(\pecos) (thresholds $\epsilon=0.1$, beam-size $b=10$) \\
		\midrule
		Unigrams (3M) 													& 38.40 & 52.49	& 54.86	&  \best{25.7} &   \best{90.0} \\
		Unigrams $+$ Bigrams (3M) 										& 39.66	& 54.93	& 57.70	&  31.5	&  212.0 \\
		Unigrams $+$ Bigrams (5.6M) 									& 40.71	& 56.09	& 58.88	&  67.9	&  232.0 \\
		Unigrams $+$ Bigrams $+$ Char Trigrams (4.2M) 					& \best{42.86}	& \best{57.78}	& \best{60.30}	&  42.5	&  295.0 \\
		%Unigrams $+$ Char Trigrams (1.2M) 								& 39.44	& 53.60	& 55.88	&  29.3	&  185.0 \\
		\midrule
		\dssm~\cite{nigam2019semantic} $+$ \hnsw~\cite{malkov2020hnsw}	& 18.03 & 30.22 & 36.19 & 263.5 &  129.0 \\
		\dssm~\cite{nigam2019semantic} $+$ exact \knn					& 18.37 & 30.72 & 36.79 & 260.0 &  192.0 \\
		\midrule
		\bmtf~\cite{robertson2009probabilistic}							& 11.62 & 18.35 & 21.72 &  &  \\
		\bottomrule
	\end{tabular}%
	}
	\caption{
		Comparing \xrlinear(\pecos) with different $n$-gram features to representative semantic matching algorithms,
		where the output space is $100$ million products.
		The results of \xrlinear are obtained by weight pruning threshold $\epsilon=0.1$ and beam search size $b=10$.
		Training time is measured in hours (hr) and model size on disk is measured in Gigabytes (GB).
		The model size of \dssm includes the encoders and storing the product embeddings for retrieval.
		The training time of \dssm model is measured on Nvidia V100 GPUs, while all other
		methods are benchmarked on Intel CPUs.
	}
	%\vspace{-1em}
	\label{tb:main-results}
\end{table*}

\subsection{Recall and Inference Latency Trade-off}
\label{sec:exp-latency}

Depending on the computational environment and resources,
semantic product search systems can have various constraints
such as model size on disk, real-time inference memory, and
most importantly, the real-time inference latency, measured
in milliseconds per query (i.e., ms/q)
under the \textit{single-thread single-CPU setup}.
In this subsection, we dive deep to study the trade-off
between Recall$@100$ and inference latency
by varying two \xrlinear hyper-parameters:
1) weight pruning threshold $\epsilon$; 2) inference beam search size $b$.

\begin{table*}[!ht]
	\centering
	\resizebox{0.975\textwidth}{!}{%
	\begin{tabular}{l|ccc|rrrr}
		\toprule
		Pruning threshold ($\epsilon$) & Recall$@10$ & Recall$@50$ & Recall$@100$ & \#Parameters & Disk Space (GBs) & Memory (GBs) & Latency (ms/q) \\
		\midrule
		$\epsilon=0.1$		& 42.86 & 57.78 & 60.30 & 26,994M & 295 & 258 & 1.2046 \\
		$\epsilon=0.2$		& 42.42	& 57.22	& 59.81	& 15,419M & 168 & 166 & 1.0176 \\
		$\epsilon=0.3$		& 41.35	& 55.98	& 58.52	&  9,957M & 109 & 118 & 0.8834 \\
		$\epsilon=0.35$		& 40.50	& 55.09	& 57.63	&  8,182M &  90 & 102 & 0.8551 \\
		$\epsilon=0.4$		& 39.41	& 53.86	& 56.35	&  6,784M &  75 &  88 & 0.8330 \\
		$\epsilon=0.45$		& 38.08	& 52.37	& 54.85	&  5,657M &  62 &  76 & 0.8138 \\
		\bottomrule
	\end{tabular}%
	}
	\caption{
		Weight pruning of the best \xrlinear(\pecos) with various thresholds $\epsilon$.
		The model memory is measured by the stable memory consumption when
		processing input queries sequentially in a single-thread real-time inference setup.
		The inference latency includes the time for featurizating query tokens into a vector
		as well as the prediction by \xrlinear model.
	}
	%\vspace{-1em}
	\label{tb:weight-pruning}
\end{table*}

\begin{table*}[!ht]
	\centering
	\begin{minipage}{0.5\textwidth}
		\centering
		\begin{tabular}{l|ccr}
			\toprule
			Beam Size ($b$) & Recall$@100$ & Latency (ms/q) & Throughput (q/s)\\
			\midrule
			$b=1$			& 20.95 & 0.1628 & 6,140.97 \\
			$b=5$			& 49.23	& 0.4963 & 2,014.82 \\
			$b=10$			& 57.63 & 0.8551 & 1,169.39 \\
			$b=15$			& 60.94 & 1.2466 &   802.20 \\
			$b=20$			& 62.72 & 1.6415 &   609.19 \\
			$b=25$			& 63.78 & 2.0080 &   498.00 \\
			$b=30$			& 64.46 & 2.4191 &   413.37 \\
			$b=50$			& 65.74 & 3.4756 &   287.72 \\
			$b=75$			& 66.17 & 4.7759 &   209.38 \\
			$b=100$			& 66.30 & 6.0366 &   165.66 \\
			\bottomrule
		\end{tabular}
	 	\caption{
			The tradeoff between retrieval performance and inference latency
			of \xrlinear(\pecos) with the threshold $\epsilon=0.35$,
			which can be flexibly controlled by the beam size $b$.
		}
		\label{tb:beam-size}
	\end{minipage}
	\hfill
	\begin{minipage}{0.4\textwidth}
		\centering
		\includegraphics[width=\textwidth]{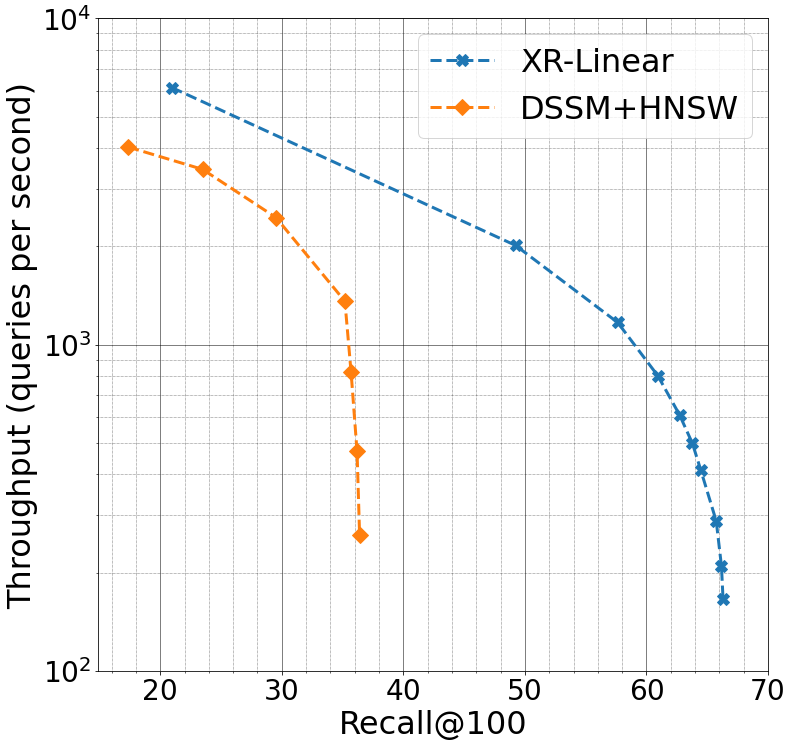}
		\captionof{figure}{
			Throughput-Recall Trade-off comparison between \xrlinear(\pecos) and \dssm$+$\hnsw.
			The curve of the latter method is obtained by sweeping the \hnsw inference
			hyper-parameters $efS=\{10, 50, 100, 250, 500\}$.
		}
		\label{fig:throughput-recall}
	\end{minipage}
\end{table*}

\paragraph{\bf Weight Pruning}
As discussed in Section~\ref{sec:model-inference},
we conduct element-wise weight pruning on the parameters
of \xrlinear model trained on Unigram$+$Bigrams$+$Char Trigrams ($4.2M$) features.
We experiment with the thresholds $\epsilon=\{0.1, 0.2, 0.3, 0.35, 0.4, 0.45\}$,
and the results are shown in Table~\ref{tb:weight-pruning}.

From $\epsilon=0.1$ to $\epsilon=0.35$,
the model size has a $3$x reduction (i.e., from $295$ GB to $90$ GB),
having the same model size as the \xrlinear model trained on unigram ($3M$) features.
Crucially, the pruned \xrlinear model still enjoys a sizable
improvement over the \xrlinear model trained on unigram ($3M$) features,
where the Recall$@100$ is $57.63\%$ versus $54.86\%$.
While the real-time inference memory of \xrlinear closely follows
its model disk space, the latter has smaller impact on the inference latency.
In particular, from $\epsilon=0.1$ to $\epsilon=0.45$, the model size
has a $5$x reduction (from $295$ GB to $62$ GB),
while the inference latency reduced by only $32.5\%$ (from $1.20$ ms$/$q to $0.81$ ms$/$q).
On the other hand, the inference beam search size has a larger impact
on real-time inference latency, as we will discuss in the following paragraph.

\begin{figure*}[!ht]
	\centering
	\includegraphics[width=0.90\textwidth]{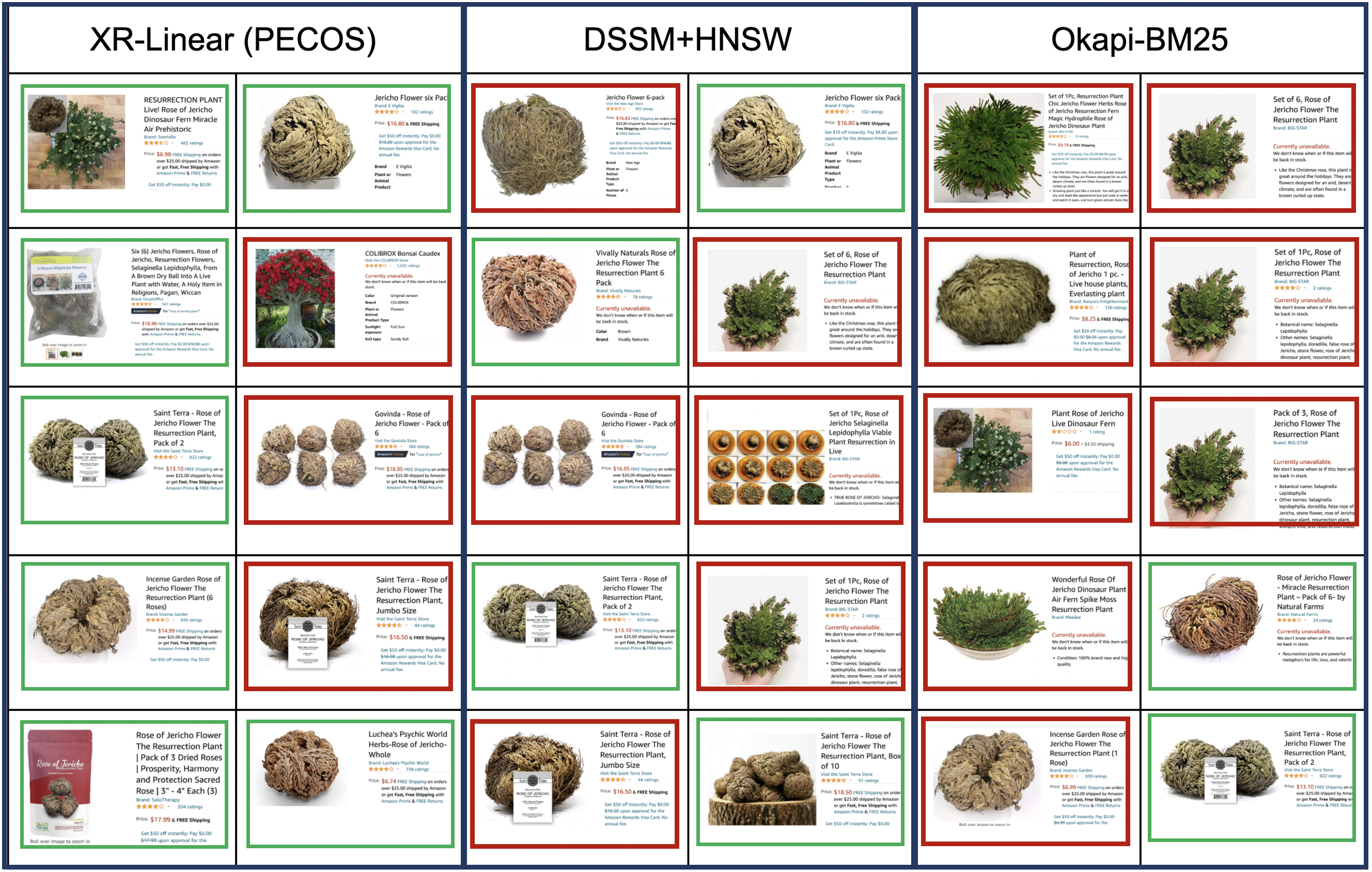}
	\vspace{-.5em}
	\caption{Side-by-side comparison of the retrieved products for \xrlinear(\pecos), \dssm$+$\hnsw, and \bmtf.
		The test query is "rose of jericho plant",
		and \textcolor{green}{green box} means there is at least one purchase while \textcolor{red}{red box} means no purchase.
	}
	\vspace{-.5em}
	\label{fig:side-by-side-v1}
\end{figure*}

\paragraph{\bf Beam Size}
We analyze how beam search size of \xrlinear prediction
affects the Recall$@100$ and inference latency.
The results are presented in Table~\ref{tb:beam-size}.
We also compare the throughput (i.e., inverse of latency, higher the better)
versus Recall$@100$ for \xlinear and \dssm $+$ \hnsw,
as shown in Figure~\ref{fig:throughput-recall}.

From beam size $b=1$ to $b=15$, we see a clear
trade-off between Recall$@100$ and latency, where
the former increases from $20.95\%$ to $60.94\%$,
at the cost of $7x$ higher latency
(from $0.16$ ms$/$q to $1.24$ ms$/$q).
Real-world product search systems typically limit
the real-time latency of matching algorithms to be
smaller than $5$ ms$/$q.
Therefore, even with $b=30$, the proposed \xrlinear
approach is still highly applicable for the real-world
online deployment setup.

In Figure~\ref{fig:throughput-recall},
we also examine the throughput-recall trade-off for the embedding-based retrieval model
\dssm~\cite{nigam2019semantic} that uses \hnsw~\cite{malkov2020hnsw} to do inference.
\xrlinear outperforms \dssm $+$ \hnsw significantly,
with higher retrieval quality and larger throughput (smaller inference latency).

In Figure~\ref{fig:side-by-side-v1}, we present the retrieved products for an example
test query "rose of jericho plant", and compare the retrieval quality of
\xrlinear(\pecos), \dssm$+$\hnsw, and \bmtf.
From the top $10$ predictions (i.e., retrieved products), we see that \xrlinear covers
more products that were purchased, and the retrieved set is more diverse, compared to the
other two baselines.

\subsection{Online Experiments}
We conducted an online A/B test to experiment different semantic matching algorithms
on a large-scale e-commerce retail product website.
The control in this experiment is the traditional lexical-based matching augmented
with candidates generated by \dssm.
The treatment is the same traditional lexical-based matching, but augmented
with candidates generated by \xrlinear instead.
After a period of time,
many key performance indicators of the treatment
are statistically significantly higher than the control,
which is consistent with our offline observations.
We leave how to combine \dssm and \xrlinear to generate better
match set for product search as future work.

\section{Discussion and Future Work}
In this work, we presented \xrlinear(\pecos), a tree-based \xmc model
to augment the match set for an online retail search system
and improved product discovery with better recall rates.
Specifically, to retrieve products from a $100$ million product catalog,
the proposed solution achieves Recall@100 of $60.9\%$ with a low latency
of $1.25$ millisecond per query (ms$/$q).
Our tree-based linear models is robust to weight pruning which can flexibly
meet different system memory or disk space requirements.

One challenge for our method is how to retrieve cold-start products
with no training signal (e.g., no clicks nor purchases).
One naive solution is to assign cold-start products
into nearest relevant leaf node clusters based on distances
between product embeddings and cluster centroids.
We then use the average of existing product weight vectors which are the nearest neighbors
of this novel product to induce the relevance score.
This is an interesting setup and we leave it for future work.
\bibliographystyle{ACM-Reference-Format}
\bibliography{midas}

%\appendix

\end{document}